\documentclass[proceedings, preprint]{rmaa}

\usepackage{paralist}

\usepackage{psfrag,color}
\SetYear{2008}
\SetConfTitle{ MAGNETIC FIELDS IN THE UNIVERSE II}

\title{Ground-state alignment of atoms and ions: New Diagnostics of Astrophysical Magnetic Field in Diffuse Medium} 

\author{
  Huirong Yan,\altaffil{CITA, 60 St George St, Toronto, M5S 3H8, Canada}
  A. Lazarian,\altaffil{University of Wisconsin-Madison, 475 Charter St., USA}}

\shortauthor{Yan \& Lazarian}
\shorttitle{Ground-state atomic alignment}

\abstract{
We discuss a new technique of studying magnetic fields in diffuse astrophysical media, e.g.
interstellar and intergalactic gas/plasma. This technique is based on the
angular momentum alignment of atoms and ions in their ground or metastable states. As the life-time
of atoms in such states is long, the alignment induced by anisotropic radiation 
is susceptible to weak magnetic fields ($1{\rm G}\gtrsim B\gtrsim0.1\mu$G). The alignment
reveals itself in terms of the polarization of the absorbed and emitted light. A variety of atoms with
fine or hyperfine splitting of the ground or metastable states exhibit the alignment and the resulting
polarization degree in some cases exceeds 20\%. We show that in the case of absorption the
polarization direction is either parallel or perpendicular to magnetic field, while more complex 
dependencies emerge for the case of emission of aligned atoms. We show that the corresponding
studies of magnetic fields can be performed with optical and UV polarimetry. A unique feature of
these studies is that they can reveal the 3D orientation of magnetic field.
 In addition, we point out that the polarization of the
radiation arising from the transitions between fine and hyperfine states of the ground level can provide yet another 
promising diagnostics of magnetic fields, including the magnetic fields in the Early Universe. We mention several
cases of interplanetary, circumstellar and interstellar magnetic fields for which the studies of magnetic fields using ground state
atomic alignment effect are promising.}
 
\addkeyword{Atomic processes}
\addkeyword{ISM: magnetic fields}
\addkeyword{polarization}

\begin{document}
% Typeset article header
\maketitle

\section{Studies of astrophysical magnetic fields}
\label{sec:intro}

Magnetic fields play extremely important roles in many astrophysical circumstances, e.g., the interstellar medium, intergalactic medium and quasars, etc.
Unfortunately, there are only a few techniques for magnetic field studies
that are available (see below). Each technique is sensitive to magnetic fields in 
particular environment. Therefore even the directions of magnetic field obtained  for
the same region of sky with different techniques differ substantially.
The simultaneous use of different techniques provides a possibility of
magnetic field tomography.

 Polarimetry of aligned dust provides a way of studying 
magnetic field direction in the diffuse interstellar medium, molecular clouds, 
circumstellar and interplanetary medium (see review by Lazarian 2007).  
Substantial progress in the understanding of grain alignment has been achieved in the last decade making the technique more reliable. The technique is, 
however, challenging to apply to low column densities, as the polarization
signal becomes too weak. It has other limitations when dealing with high 
densities (Cho \& Lazarian 2005).

Polarimetry of some molecular lines using the Goldreich-Kylafis (1982) effect has recently been shown to be a good tool for magnetic field  studies in molecular clouds (see Girart, Crutcher \& Rao 1999). However, the magnetic field direction obtained has an uncertainty of 90 degrees, which may be confusing. The 
technique is most promising for dense CO clouds.

Zeeman splitting (see Crutcher 2004) provides a good way to get magnetic
field strength, but the measurements are very time consuming and only
the strongest magnetic fields are detectable this way 
(see Heiles \& Troland 2004).

Synchrotron emission/polarization as well as Faraday rotation provide an
important means to study magnetic fields either in distinct regions with strong magnetic
fields or over wide expanse of the magnetized diffuse media (Haverkorn 2005).

Here we discuss a new promising technique to study magnetic fields in diffuse medium. The recent development of this field can be found in Yan \& Lazarian (2006, 2007, 2008 henceforth YLa,b,c). 
The technique employs spectral-polarimetry and makes use of
the ability of atoms and ions to be aligned {\it in their ground state} by the external anisotropic radiation.
The aligned atoms interact with the astrophysical magnetic fields to get realigned. As the life time
of the ground state is long, even weak magnetic fields can be detected this way. The requirement for the
alignment in the ground state is the fine or hyperfine splitting of the ground state. The latter is true for
many species present in diffuse astrophysical environments.
Henceforth, we shall not
 distinguish atoms and ions and use word ``atoms'' dealing  with both species.
This technique can be used for 
interstellar\footnote{Here interstellar is understood in a 
general sense, which, for instance,
includes refection nebulae.}, and
 intergalactic studies as well as for studies of magnetic fields in
QSOs and other astrophysical objects.  

The effect of ground-state atomic alignment is based on the well known physics. 
In fact, it has been known that atoms can be aligned through interactions with the anisotropic flux
of resonance emission (see review Happer 1972 and references therein). Alignment is  understood here in terms of orientation
of the angular momentum vector
$\bf J$, if we use the language of classical mechanics. In quantum
terms  this means a difference in the population of sublevels corresponding to
projections of angular momentum to the quantization axis. Whenever this does not cause confusion, we use "atomic alignment" instead of
more precise "ground-state atomic alignment".

It is worth mentioning that atomic alignment was
studied in laboratory in relation with early-day maser 
research\footnote{Our studies in YLa) revealed that the mathematical treatment of the effect was not adequate, however.} (see Hawkins 1955). This effect was discussed in the interstellar medium context 
by Varshalovich (1968) for an atom with a hyperfine splitting. Varshalovich (1971) pointed out that atomic alignment enables us to detect the direction of magnetic fields in the interstellar medium, but did not provide a necessary quantitative study.  

A case of emission of an idealized fine structure atom subject to a magnetic field and a beam of pumping radiation was conducted in Landolfi \& Landi Degl'Innocenti (1986). However, in that case, an idealized 
two-level atom was considered. In addition, polarization of emission from this atom  was
discussed for a very restricted geometry of observations,
namely, the magnetic field is along the line of sight and both of these directions are perpendicular to the beam of incident light. 
This made it rather difficult to use this study as a tool for practical mapping of magnetic fields in various astrophysical environments.

It should be pointed out that the atomic alignment we deal with in this paper differs from the Hanle effect that solar researcher have studied. Hanle effect is depolarization and rotation of the polarization vector of the resonance scattered lines in the presence of a magnetic field, which happens when the magnetic splitting becomes comparable to the decay rate of the excited state of an atom. The research into emission line polarimetry resulted in important
change of the views on solar chromosphere (see Landi Degl'Innocenti 1983, 1984, 1998, Stenflo \& Keller 1997, Trujillo
Bueno \& Landi Degl'Innocenti 1997, Trujillo Bueno et al. 2002). However, these studies correspond to a setting different from the one we consider here.
In this paper we concentrate on the weak field regime, in which it is 
the atoms at ground level that are repopulated due to magnetic precession, while the Hanle effect is negligible for the upper state. 
This is the case, for instance, of the interstellar medium. The polarization of absorption lines is thus more
informative. In many cases, we are in optically thin regime, so we do not need to be concerned about radiative transfer.

\section{Conditions for atomic alignment}

The basic idea of the atomic alignment is quite
simple. The alignment is caused by
the anisotropic deposition of angular momentum from photons. In typical
 astrophysical situations the radiation
flux is anisotropic (see Fig.\ref{nzplane}{\it right}). As the photon
spin is along the direction of its propagation, we expect that atoms
scattering the radiation from a light beam to be  aligned. Such an alignment happens in terms of 
the projections of angular momentum
to the direction of the incoming light. For atoms to be aligned,
 their ground state should have non-zero angular momentum. Therefore fine (or hyperfine) structure is necessary to enable various projection of atomic angular momentum to exist in their ground state. 

\subsection{Basics of atomic alignment}
\begin{figure*}[!t]
\includegraphics[%
  width=0.7\textwidth,
  height=0.25\textheight]{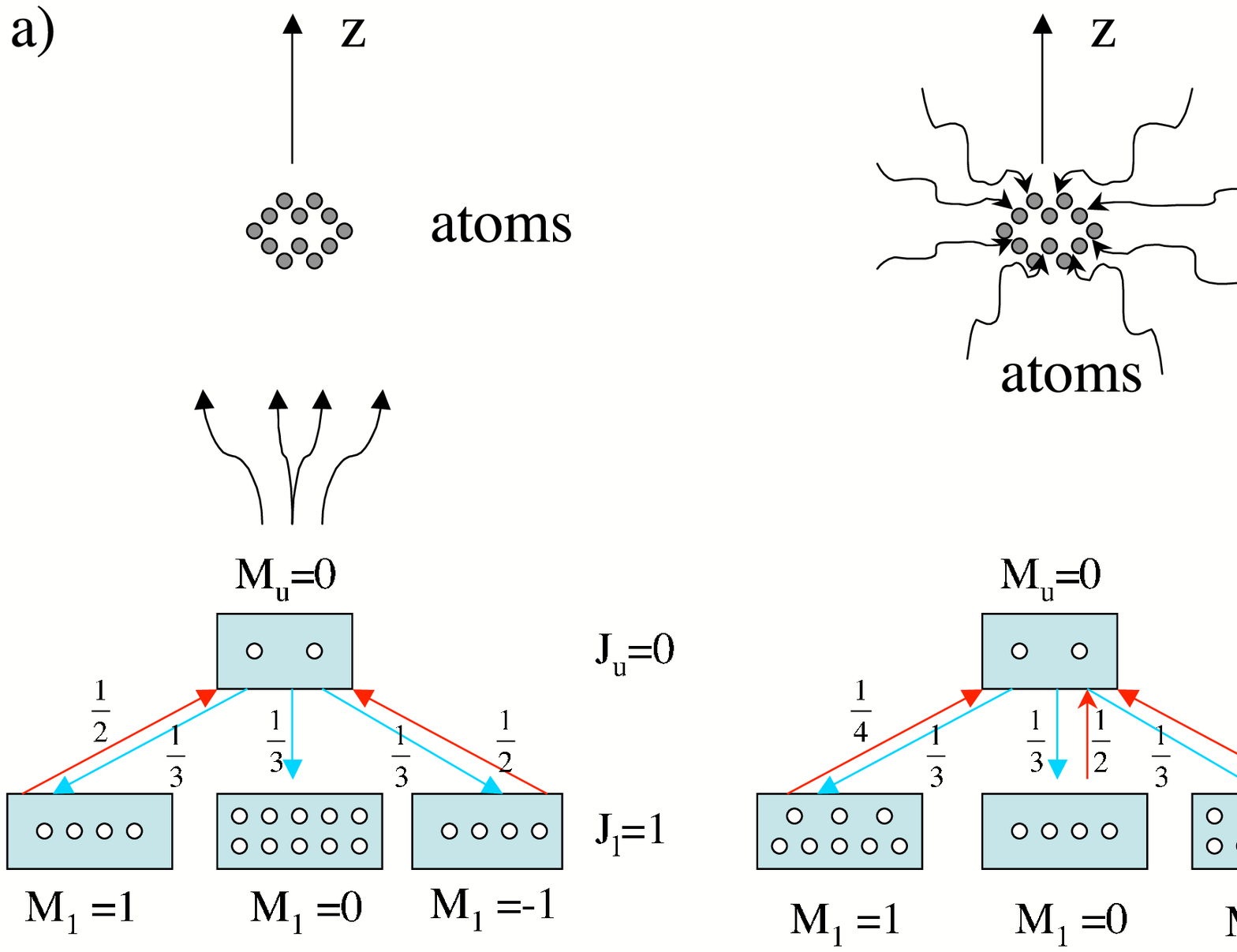}
\includegraphics[%
  width=0.25\textwidth,
  height=0.25\textheight]{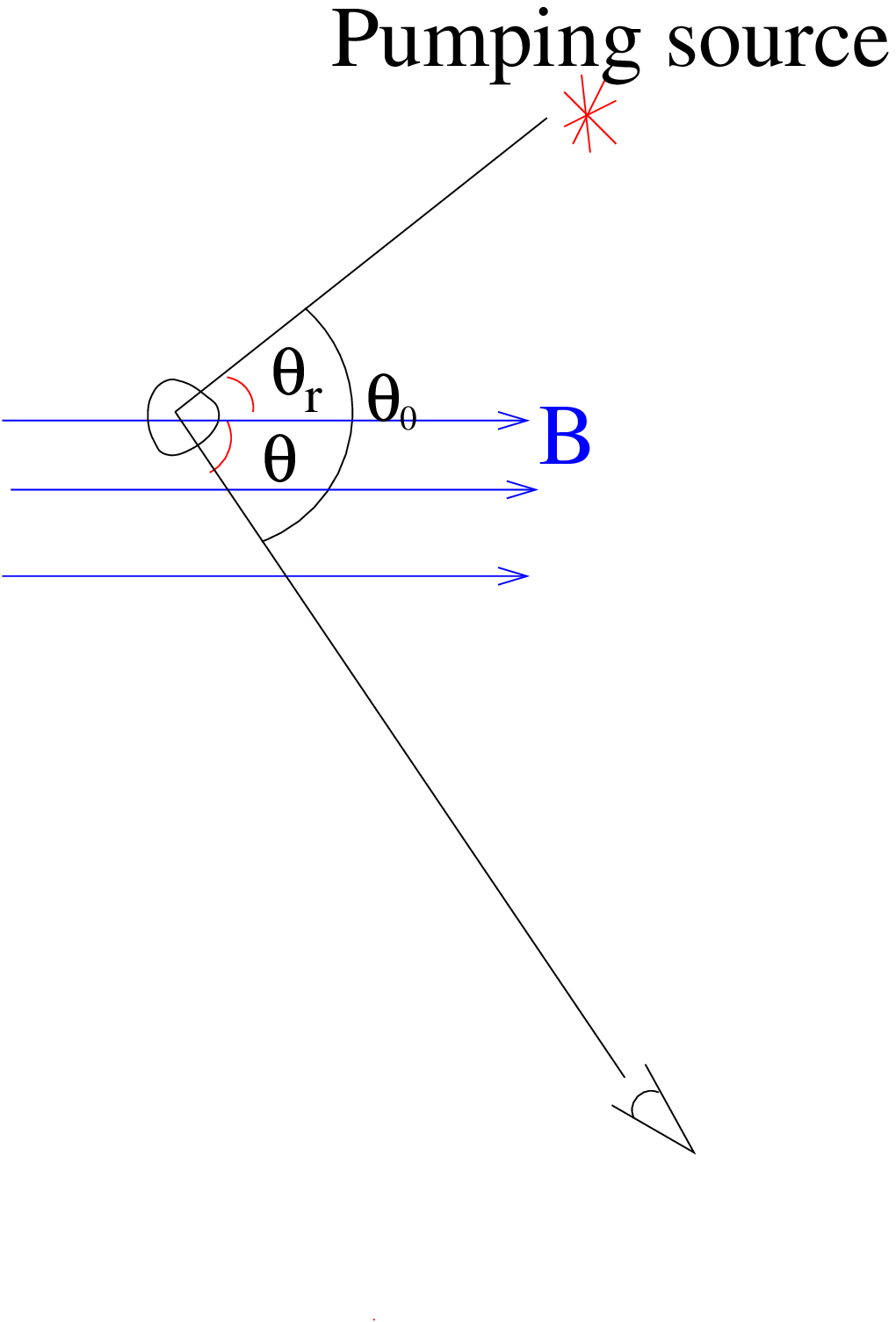}
\caption{{\it Left}: A toy model to illustrate how atoms are aligned by anisotropic light.
Atoms accumulate in the ground sublevel $M=0$ as radiation removes atoms from the ground states $M=1$ and $M=-1$; {\it right}: Typical astrophysical environment where the ground-state atomic alignment can happen. A pumping source deposits angular momentum to atoms in the direction of radiation and causes differential occupations on their ground states. In a magnetized medium where the Larmor precession rate $\nu_L$ is larger than the photon arrival rate $\tau_R^{-1}$, however, atoms are realigned with respect to magnetic field. Atomic alignment is then determined by $\theta_r$, the angle between the magnetic field and the pumping source. The polarization of scattered line also depends on the direction of line of sight, $\theta$ and $\theta_0$. (From YLc)}
\label{nzplane}
\end{figure*}

Let us discuss a toy model that provides an intuitive  insight into  the 
physics of atomic alignment.
Consider an atom with its ground state corresponding to the total angular momentum
$I=1$ and the upper state corresponding to the angular momentum $I=0$ (Varshalovich 1971).
If the projection of the angular momentum to the direction of the
incident resonance photon beam is $M$, the upper state $M$ can
have values $-1$, $0$, and $1$, while for the upper state M=0 (see Fig.\ref{nzplane}{\it left}). 
The unpolarized
beam contains an equal number of left and right circularly polarized
photons whose projections on the beam direction are 1 and -1. Thus
absorption of the photons will induce transitions from the $M=-1$ and
$M=1$ sublevels. However, the decay from the upper state populates all
the three sublevels on ground state. As the result the atoms accumulate in the $M=0$ ground
sublevel from which no excitations are possible. Accordingly, the optical
properties of the media (e.g. absorption) would change.

The above toy model can also exemplify the
 role of collisions and magnetic field. Without collisions one may expect 
that all atoms
reside eventually at the sublevel of $M=0$. Collisions, however, redistribute
atoms to different sublevels. Nevertheless, as disalignment of the ground
state requires spin flips, it is less efficient than one might naively
imagine (Hawkins 1955). The reduced sensitivity of aligned
atoms to disorienting collisions makes the effect important for various
astrophysical environments.

Owing to the precession, the atoms with different projections of angular momentum will be mixed up. 
Magnetic mixing happens if the angular
momentum precession rate  is higher than the rate of the
excitation from the ground state, which is true for many astrophysical conditions, e.g., interplanetary medium, ISM, intergalactic medium, etc. 
As the result, angular momentum is redistributed among the atoms, and the alignment is altered according to the angle between the magnetic field and radiation field $\theta_r$ (see Fig.\ref{nzplane}{\em right}). This is the {\em classical} picture. 

In {\em quantum} picture, if magnetic precession is dominant, then the natural quantization axis will be the magnetic field, which in general is different from the symmetry axis of the radiation. The radiative pumping is to be seen coming from different directions according to the angle between the magnetic field and radiation field $\theta_r$, which results in different alignment.

The classical theory can give a qualitative interpretation which shall be utilized in this paper to provide an intuitive picture. Particularly for emission lines, both atoms and the radiation have to be described by the density matrices in order to obtain quantitative results. This is because there is coherence among different magnetic sublevels on the upper state.

All in all, in order to be aligned, first, atoms should have enough 
degrees of freedom: namely, the quantum angular momentum number must be 
$\ge 1$. Second, the incident flux must be anisotropic. 
Moreover, the collisional rate should not be too high. While the latter
requires special laboratory conditions, it is applicable to many astrophysical environments such as the outer
layers of stellar atmospheres, the interplanetary,
interstellar, and intergalactic medium. 

Many species satisfy the above conditions and can be aligned. The corresponding lines (including both absorption and emission) lines can be used as the diagnostics. A number of lines with the maximum degree of polarization have been provided in YLa,b,c. Table \ref{species} lists a couple of lines we recently calculated as suggested by a number of observers.

\begin{table*}
\tablecols{6}
\caption{{\tiny The polarization of two emission lines.}}
\begin{tabular}{cccccc}
\toprule
Species&Nuclear spin&Lower level&Upper level&Wavelength&$P_{max}$\\
\midrule
Al II&5/2&$1S_0$&$1P^o_1$&8643\AA&20\%\\
C I&0&$3P^o_0$&$3P^o_1$&$610\micron$&20\%\\
\bottomrule
\end{tabular}
\label{species}
\end{table*} 

\subsection{Relevant Timescales}
Various species with fine structure 
can be aligned. A number of selected transitions  that can be used
for studies of magnetic fields are listed in YLa,b,c and table~\ref{species}. Why and how are these lines chosen?  We gathered all of the prominent interstellar and intergalactic lines (Morton 1975, Savage et al. 2005), from which we take only alignable lines, namely, lines with ground angular momentum number $J_g (or F_g)\geq1$. The
number of prospective transitions increases considerably if we add
QSO lines. In fact, many of the species listed in the Table 1 in Verner, Barthel, \& Tytler (1994) are alignable and observable from the ground because of the cosmological
redshifts. In this paper we do not consider such transitions. 

In terms of practical magnetic field studies, the variety of available species is important in many aspects. One of them is a possibility of
getting additional information about environments. Let us illustrate this by considering
the various rates (see Table \ref{difftime}) involved. Those 
are 1) the rate of the Larmor precession, $\nu_L$, 2) the rate of the optical pumping, $\tau_R^{-1}$, 3) the rate of collisional randomization, $\tau_c^{-1}$,
4) the rate of the transition within ground state, $\tau^{-1}_T$.  
In many cases $\nu_L>\tau_R^{-1}>\tau_c^{-1}, \tau_T^{-1}$. 
Other relations are possible, however. If $\tau_T^{-1}>\tau_R^{-1}$, the transitions within the sublevels of ground state need to be taken into account and relative distribution among them will be modified (see YLa,c). Since emission is spherically symmetric, the angular momentum in the atomic system is preserved and thus alignment persists in this case. In the case $\nu_L<\tau_R^{-1}$, the magnetic field does not affect the atomic occupations and atoms are aligned with respect to the direction of radiation. From the expressions in Table~\ref{difftime}, we see, for instance, that magnetic field can realign CII only at a distance $r\gtrsim7.7$Au from an O star if the magnetic field strength $\sim 5\mu$G.

If the Larmor precession rate $\nu_L$ is comparable to any of the other rates,
the atomic line polarization becomes sensitive to the strength of the magnetic field. In these situations, it is possible to get information about the {\it magnitude} of magnetic
field. 

\begin{table*}[!t]
\tablecols{4}
\begin{tabular}{cccc}
\toprule
$\nu_L$(s$^{-1}$)&$\tau_R^{-1}$(s$^{-1}$)&$\tau_T^{-1}$(s$^{-1}$)&$\tau_c^{-1}$(s$^{-1}$)\\
\midrule
$\frac{eB}{m_ec}$&$B_{J_lJ_u}I$&$A_m$&max($f_{kj},f_{sf}$) \\
$88(B/5\mu$ G)& $7.4\times 10^{5}\left(\frac{R_*}{r}\right)^2$&2.3$\times 10^{-6}$&$6.4\left(\frac{n_e}{0.1{\rm cm}^{-3}}\sqrt{\frac{8000{\rm K}}{T}}\right)\times 10^{-9}$\\
\bottomrule
\end{tabular}
\caption{{\tiny Relevant rates for atomic alignment. $A_m$ is the magnetic dipole emission rate for transitions among J levels of the ground state of an atom. $f_{kj}$ is the inelastic collisional transition rates within ground state due to collisions with electrons or hydrogens, and $f_{sp}$ is the spin flip rate due to Van der Waals collisions. In the last row, example values for C II are given. $\tau_R^{-1}$ is calculated for an O type star, where $R_*$ is the radius of the star radius and r is the distance to the star. (From YLa)}}
\label{difftime}
\end{table*}

\begin{figure}[!t]
  \includegraphics[width=\columnwidth,height=0.25\textheight]{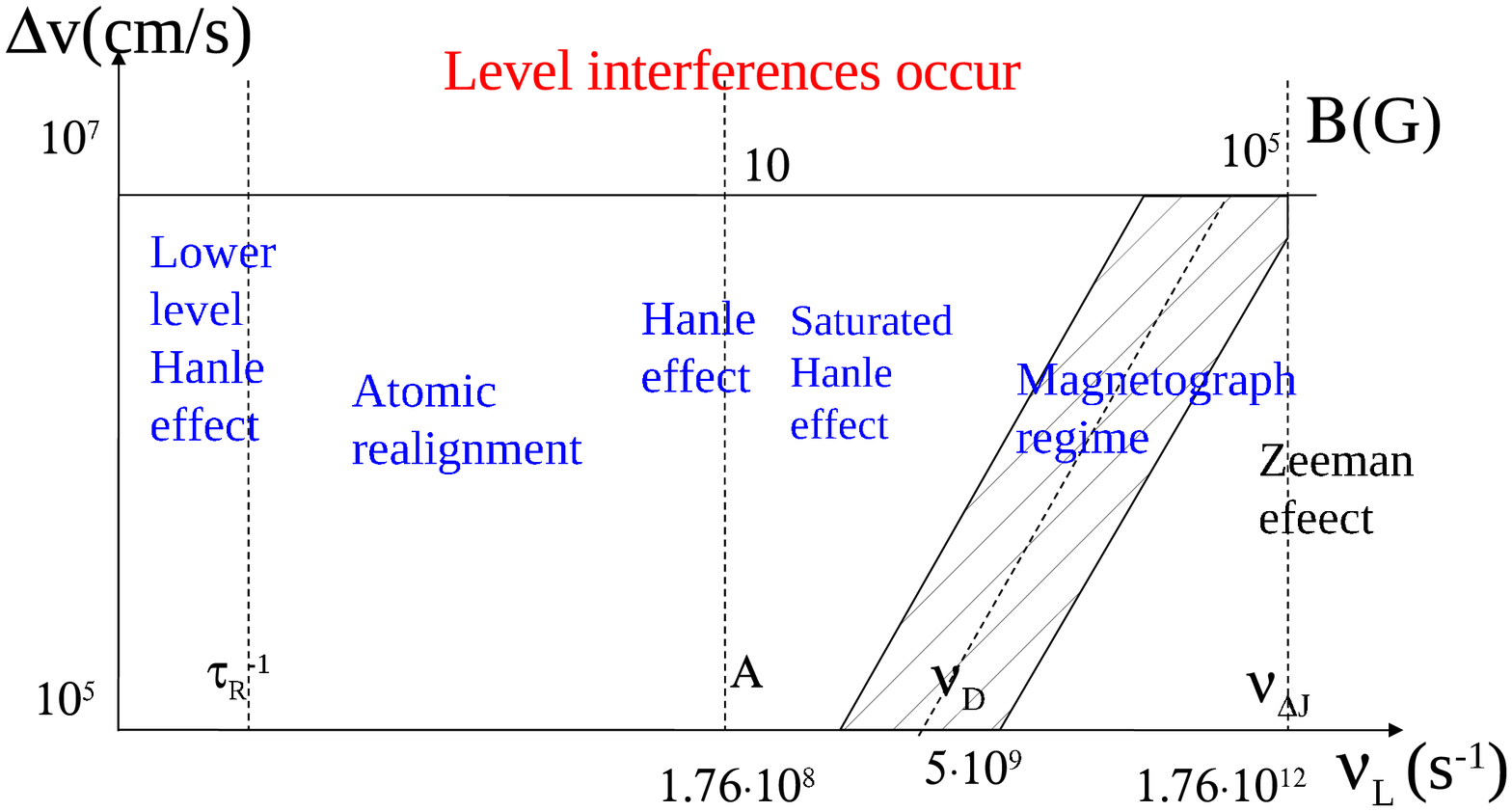}
  \caption{ Different regimes divided according to the strength of magnetic field and the Doppler line width. Atomic realignment is applicable to weak field ($<1G$) in diffuse medium. Level interferences are negligible unless the medium is substantially turbulent ($\delta v\gtrsim$ 100km/s) and the corresponding Doppler line width becomes comparable to the fine level splitting $\nu_{\Delta J}$. For strong magnetic field, Zeeman effect dominates. When magnetic splitting becomes comparable to the Doppler width, $\sigma$ and $\pi$ components (note: we remind the reader that $\sigma$ is the circular polarization and $\pi$ represents the linear polarization.) can still distinguish themselves through polarization, this is the magnetograph regime; Hanle effect is dominant if Larmor period is comparable to the lifetime of excited level $\nu_L^{-1}\sim A^{-1}$; similarly, for ground Hanle effect, it requires Larmor splitting to be of the order of photon pumping rate; for weak magnetic field ($<1G$) in diffuse medium, however, atomic alignment is the main effect provided that $\nu_L=17.6(B/\mu G){\rm s}^{-1}>\tau_R^{-1}$. (From YLc)}
  \label{regimes}
\end{figure}

Fig.\ref{regimes} illustrates the regime of magnetic field strength where atomic realignment applies. Atoms are aligned by the anisotropic radiation at a rate of $\tau_R^{-1}$. Magnetic precession will realign the atoms in their ground state if the Larmor precession rate $\nu_L>\tau_R^{-1}$. In contrast, if the magnetic field gets stronger so that Larmor frequency becomes comparable to the line-width of the upper level, the upper level occupation, especially coherence is modified directly by magnetic field, this is the domain of Hanle effect, which has been extensively discussed for studies of solar magnetic field (see Landi Degl'Innocenti 2004 and references therein). When the magnetic splitting becomes comparable to the Doppler line width $\nu_D$, polarization appears, this is the ``magnetograph regime" (Landi Degl'Innocenti 1983). For magnetic splitting $\nu_L\gg \nu_D$, the energy separation is enough to be resolved, and the magnetic field can be deduced directly from line splitting in this case. If the medium is strongly turbulent with $\delta v\sim 100$km/s (so that the Doppler line width is comparable to the level separations $\nu_D\sim \nu_{\Delta J}$), interferences occur among these levels and should be taken into account.

Long-lived alignable metastable states that are present for
some atomic species between upper and lower states may act as
proxies of ground states.  Absorptions from these metastable levels
 can also be used as diagnostics for magnetic field therefore.

\section{Applications: synthetic observation of various objects-from solar system to early universe} 

\subsection{Magnetic field traced by the Sodium in comet wake}
\label{cometwake}

\begin{figure*}[!t]
\includegraphics[width=0.45\textwidth,height=0.22\textheight]{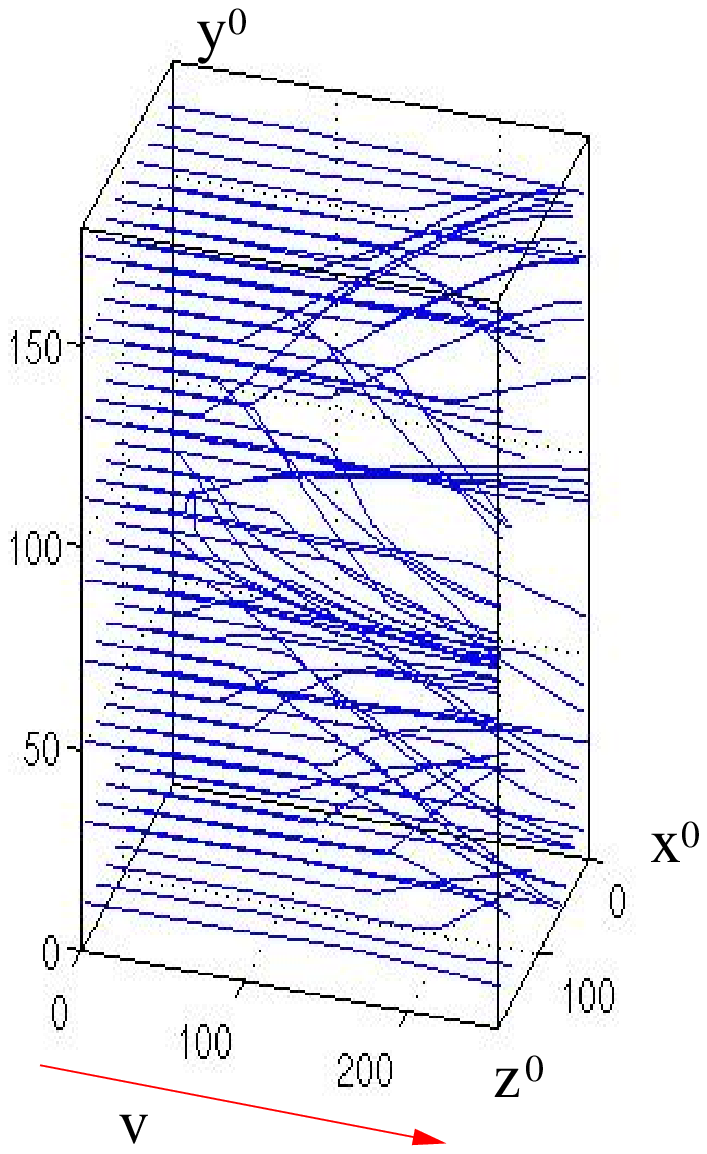}
\includegraphics[width=0.45\textwidth,height=0.22\textheight]{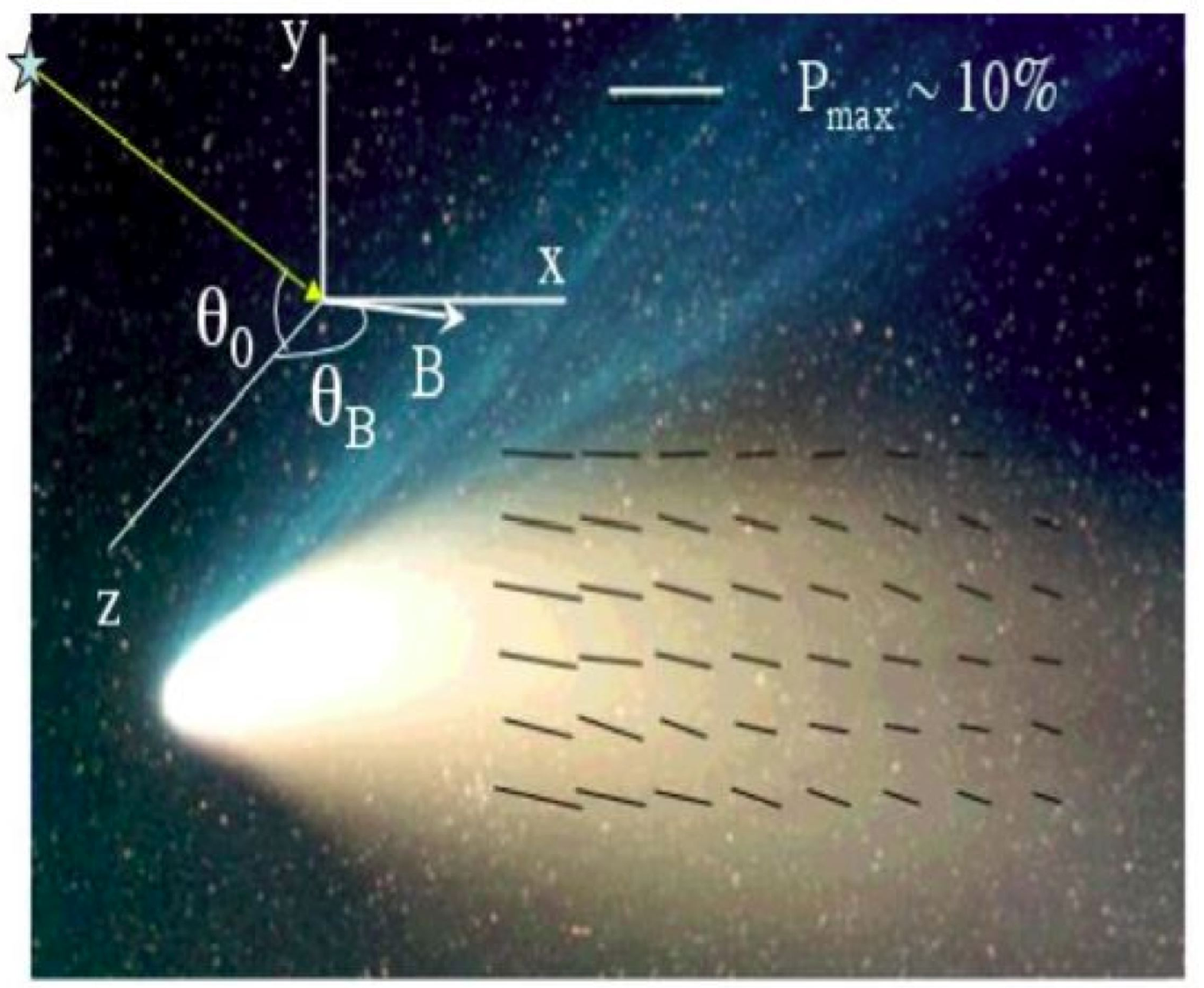}
\caption{{\it Left}: simulated magnetic field distribution in the comet wake; {\it right}: the map of a synthetic observation of the polarization of Na D2 emission from the comet wake. (From YLb)}
\label{comet}
\end{figure*}

As an illustration, we discuss here a synthetic observation of a comet wake. 
Though the abundance of sodium in comets is very low, its high efficiency of
 scattering sunlight makes it a good tracer 
(Thomas 1992). It was suggested by Cremonese \& Fulle (1999) there are two categories of sodium tails. Apart from the {\it diffuse} sodium tail superimposed on dust tail, there is also a third {\it narrow} tail composed of only neutral sodium and well separated from dust and ion tails. This neutral sodium tail is characterized by fast moving atoms from a source inside the nuclear region and accelerated by radiation pressure through resonant D line scattering. While for the diffuse tail, sodium are considered to be released in situ by dust, it is less clear for the second case. Possibly the fast narrow tail may also originate from the rapidly fragmenting dust in the inner coma (Cremonese et al. 2002). 

The gaseous sodium atoms in the comet tail acquires not only momentum, but also angular momentum from the solar radiation, i.e. they are aligned.  Distant from comets, the Sun can be 
considered a point source. As shown in Fig.\ref{comet}, the geometry of the 
scattering is well defined, i.e., the scattering angle $\theta_0$ is known. 
The polarization of the sodium emission thus provides an exclusive 
information of the magnetic field in the comet wake.  Embedded in Solar 
wind, the magnetic field is turbulent in a comet wake. We take a 
data cube (Fig.\ref{comet}) from MHD simulations of a comet wake. Depending on its 
direction, the embedded magnetic field alter the degree of 
alignment and therefore polarization of the light scattered by the aligned atoms. Therefore, fluctuations in the linear polarization are expected from 
such a turbulent field. The calculation is done for the equilibrium case. If otherwise, the result for degree of polarization will be slightly different ($\lesssim \%10$) depending on the number of scattering events experienced by atoms (see YLb). The direction of polarization, nevertheless, should be the same as in the equilibrium cases. Except from polarization, intensity can also be used as a diagnostic. By comparing observations with it, we can determine whether magnetic field exists and their directions. For interplanetary studies,
one can investigate not only spatial, but also temporal variations
of magnetic fields. Since alignment happens at a time scale $\tau_R$, magnetic field variations on this time scale will be reflected. This can allow cost effective way of studying interplanetary magnetic turbulence at different scales.

\subsection{Magnetic field in the circumstellar region}

\begin{figure}
\includegraphics[width=.9\columnwidth,height=0.25\textheight]{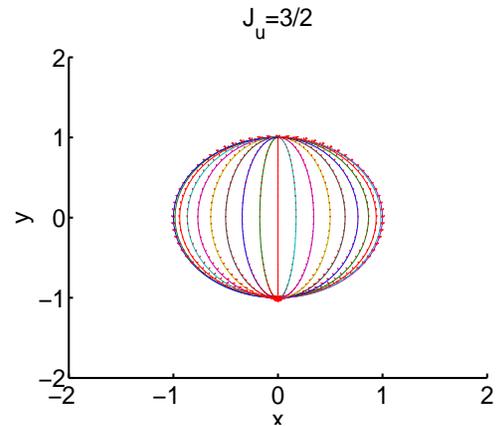}
\caption{Polarization vectors of OI emission in a circumstellar region with uniform magnetic field. The inclination of magnetic field is 30 degree from the light of sight. The magnetic field is in the y direction in the plane of sky.}
\label{circum}
\end{figure}

Aligned species can also be used to diagnose the magnetic field in the circumstellar region. We performed calculations for a circumstellar envelope with uniform magnetic field. Fig.\ref{circum} shows the corresponding polarization of scattered light from aligned O I atoms.

\subsection{Magnetic field in the epoch of reionization?}
\label{mag_dipole}
The issue of magnetic field at the epoch of reionization is a subject of controversies. The fact that the levels of an atomic ground state can be aligned through anisotropic pumping suggest us a possibility of using atomic alignment to diagnose whether magnetic field exists at that early epoch. 

\begin{figure} 
\includegraphics[width=\columnwidth,height=0.25\textheight]{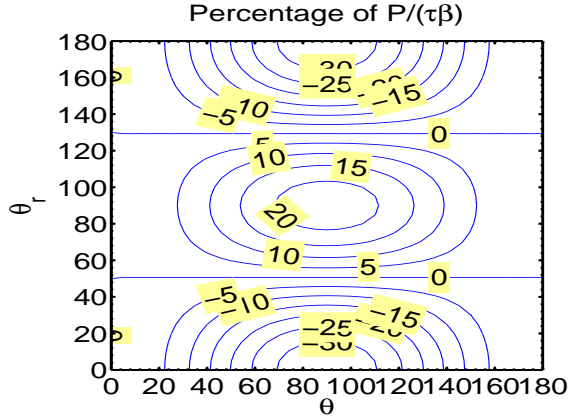}
\caption{The contour plots of equal percentages of the polarization $P/(\tau\beta)$. $\theta_r,\, \theta$ are respectively the angles of the incident radiation and l.o.s. from the magnetic field. (From YLb).}
\label{OImicroforB}
\end{figure}

In the case of nonzero magnetic field, the density matrices are determined by $\theta_r$, the angle between magnetic field and the symmetry axis of the magnetic field, as well as the parameter $\beta=BI_\nu/B_mI_m$, the ratio of UV excitation rate to the CMB excitation rate. The degree of polarization is proportional to $\beta$. In Fig.\ref{OImicroforB}, we show the dependence of the ratios $P/(\tau\beta)$ on $\theta_r$ and $\theta$, where $\tau$ is the optical depth. The line is polarized either parallel ($P>0$) or perpendicular ($P<0$) to the magnetic field. The switch between the two cases happen at $\theta_r=\theta_V=54.7^o, 180-54.7^o$, which is a common feature of polarization from aligned level (see YLa,b for detailed discussions). 

We discussed pumping of hyperfine lines [H I] 21 cm and [N V] 70.7 mm in YLb and fine line [O I] $63.2\micron$ in YLc. The most recent result on [CI] $610\micron$ is shown in Table~\ref{species}. Certainly this effect widely exists in all atoms with some structure on ground state, e.g., Na I, K I, fine structure lines, [C II], [Si II], [N II], [N III], [O II], [O III],[S II], [S III], [S IV], [Fe II], etc (see Table~4.1 in Lequeux 2005). Many atomic radio lines are affected in the same way and they can be utilized to study the physical conditions, especially in the early universe: abundances, the extent of reionization through the anisotropy (or localization) of the optical pumping sources, and {\em magnetic fields}, etc.

\section{Discussion}

\subsection{Range of applicability}

An incomplete list of objects where effects of alignment should be accounted for arises from our studies, which include this paper, as well as YLa,b. These include diffuse medium in the early Universe, quasars, AGNs, reflection nebulae, high and low density ISM, circumstellar regions, accretion disks and comets. One can easily add more astrophysical objects to this list. For instance, Io sodium tail can be studied the same way as sodium tail of comets. 

In general, in all environment when optical pumping is fast compared with the
collisional processes we expect to see effects of atomic alignment and magnetic realignment of atoms. The wide variety of atoms with fine and hyperfine structure of levels ensures multiple ways that the information can be obtained. Comparing information obtained through different species one can get deep insights into the physics of different astrophysical objects. If the implications of atomic alignment influenced the understanding of particular features of the Solar spectrum, then the studies of atomic alignment in a diffuse
astrophysical media can provide much deeper and yet unforeseen changes in our understanding of a wide variety of physical processes.

As the resolution and sensitivity
of telescopes increases, atomic alignment will be capable to 
probe the finer structure of astrophysical magnetic fields including those in the halo of accretion disks, stellar winds etc. Space-based polarimetry should provide a wide variety of species to study magnetic fields with.

\subsection{Polarization of absorption lines}
The direction of polarization has the same pattern, namely, either $\parallel$ or $\perp$ to the magnetic field on the plane of sky (YLa). In fact, this should be applicable to 
all absorption lines (including molecular lines) regardless of their different structures. 

This fact is very useful in practice. It means that even we do not have an exact prediction and precise measurement of the degree of polarization of the absorption lines. We can have a 2D mapping of magnetic field on the plane of sky within an accuracy of $90^o$ once we observe their direction of polarizations. In this sense, it has some similarity with the Goldreich-Kylafis effect although it deals with radio emission lines.

For absorption lines, there is inevitably dilution along line of sight, which adds another dependence on the ratio of alignable column density and total column density $N_a/N_{tot}$. For different species, this ratio is different. The ratio should be close to 1 for highly ionized species which only exist near radiation sources. The same is true
 for the absorption from metastable state (see Paper I). 
Combining different species (with different $N_a/N_{tot}$), it is possible to acquire a tomography of the magnetic field {\it in situ}. 

\subsection{Polarization of emission lines}

Compared to absorption lines, emission lines are more localized and therefore the dilution along the line of sight can be neglected. The disadvantage of emission lines compared to absorption lines is that the direction of the polarization of emission lines has a complex dependence on the direction of the magnetic field 
and the illusion light. Therefore, the use of
emission lines is more advantageous when combined with other measurements. 

We considered the situation that atoms are subjected to the 
flow of photon that excite transitions at the rate $\tau^{-1}_R=B_{lu}{\bar J}^0_0$ which is
smaller than the Larmor precession rate $\nu_L$, but larger than the rate of disalignment due to collisions $\tau^{-1}_c$. For the cold
gas with 30 hydrogen atoms per cubic cm, the characteristic range over which the atoms can be aligned by an O star is $\lesssim 15$pc for HI due to spin exchange collisions (with rate $C_{10}/n_H=3.3\times 10^{-10}{\rm cm}^3{\rm s}^{-1}$). 

\subsection{Radio Emission Lines Influenced by Optical Pumping}

{\it Anisotropic} optical pumping also influences the magnetic
dipole transitions between the sublevels of the ground state. We briefly discussed this in \S\ref{mag_dipole} and YLb,c. We showed that the emission arising from such atoms is polarized, which provides a new way of studying magnetic fields. Apart from apparent galactic and extragalactic applications, this may be an interesting process to
study magnetic fields at the epoch of reionization, which hopefully will be available with the instruments are currently under construction. Atomic alignment has some similarity to
 Goldreich-Kylafis effect, which also measures magnetic field through 
magnetic mixing (Goldreich \& Kylafis 1982; Girart, Crutcher, \& Rao 1999). 
However, Goldreich-Kylafis effect is due to the radio pumping.
Atomic alignment, on the other hand, happens with ground states 
as a result of optical and UV pumpings.  

\subsection{Comparison to using aligned grains}

In our studies we were focused on new ways to study magnetic field that atomic alignment of the atoms/ions with fine and hyperfine structure provides. Being alignable in their ground state, these species can be realigned in weak magnetic fields, which is extremely good news for the studies of weak magnetic fields in astrophysical diffuse gas. The latter studies are currently very limited, with polarimetry based on grain
alignment  being the most widely used technique (see Whittet 2005 and ref. therein). However, in spite of the progress of grain alignment
theory (see Lazarian 2007 for a review), the quantitative studies of magnetic fields with aligned atoms are not always possible. Atoms, unlike
dust grains, have much better defined properties, which allows, for instance, tomography of magnetic fields by using different species\footnote{As we discussed in YLa long-lived alignable metastable states that are present for
some atomic species between upper and ground states may act as
proxies of ground states. The life time of the metastable level
may determine the distance from the source over which the atoms
are aligned being on metastable level. 
Absorption from such metastable levels
 can be used as diagnostics for magnetic field in the star vicinity.}, which would be differentially aligned at different distances from the source. In addition, ions can trace magnetic fields in the environments in which 
grains cannot survive. 

Incidentally, for the grain alignment, anisotropic radiation has also been identified as a major driving agent (see Dolginov  \& Mytrophanov 1976, Draine \& Weingartner 1996, Lazarian \& Hoang 2007a). Another class of grain alignment mechanisms includes the mechanical alignment (see Gold 1952, Roberge et al. 1995, Lazarian \& Hoang 2007b), which arises from streaming of grains in respect to gas. Interestingly enough, atomic alignment due to collisions with atoms and ions is feasible as well (see Fineschi \& Landi Degl'Innocenti 1992). For instance, one may expect that atoms produced by charge exchange of directed flow of ions and atoms to be aligned. 

\subsection{Studying time variations of magnetic fields}

In fact, rather than compare advantages and disadvantages of different ways of studying magnetic fields, we would  stress the complementary nature of different ways of magnetic field studies. The subject is starved for both data and new approaches to getting the data. Note, that atomic realignment happens on the Larmor precession time, which potentially allows to
study the dynamics of fast variations of magnetic fields, e.g., related to MHD turbulence.
 
Note, that for interplanetary studies,
one can investigate not only spatial, but also temporal variations
of magnetic fields. This can allow cost effective way of studying
interplanetary magnetic turbulence at different scales.

\section{Summary}

$\bullet$ Atoms and ions with fine or hyperfine structure can be aligned in their ground state, providing a way of
getting unique information about weak magnetic fields in diffuse medium.

$\bullet$ Polarization of optical and UV absorption and emission
 lines can be used for the studies making use of the atomic alignment. 3D direction of magnetic field may
 be available, if several aligned transitions are used.

$\bullet$ Polarization of radio emission arising from decay of the sublevels of the aligned ground state level opens 
an alternative avenue of magnetic field studies. In particular, this may be a way to study magnetic fields in the Early Universe.

$\bullet$ Variations of the polarization arising from aligned atoms is found to follow closely the time variation of magnetic fields, which allow a cost
 effective way of studying interplanetary turbulence. 

\begin{acknowledgments}
We are grateful to Ken Nordsieck for valuable comments and suggestions. We thank Martin Houde for fruitful discussions and suggestions.
HY is supported by CITA and the National Science and Engineering Research Council of Canada. AL thanks the NSF funded Center for Magnetic Self-Organization in  Astrophysical and Laboratory Plasmas, NSF grants AST 0507164 and 0808118. 
\end{acknowledgments}

\end{document}